\documentclass[manuscript]{acmart}
\AtBeginDocument{%
  }
\setcopyright{none}
\renewcommand\footnotetextcopyrightpermission[1]{}
\acmSubmissionID{}
\setcopyright{none}
\renewcommand\footnotetextcopyrightpermission[1]{}

\usepackage{tabularx}
\usepackage{booktabs}
\usepackage{array}
\usepackage{cleveref}
\usepackage{tikz}
\usepackage{xspace}
\usetikzlibrary{positioning,arrows.meta}
\usepackage[most]{tcolorbox}
\usepackage[normalem]{ulem}
\usepackage{enumitem}
 
\definecolor{takeawaybg}{HTML}{FFF7ED}
\definecolor{takeawayborder}{HTML}{FB923C}
\definecolor{takeawayaccent}{HTML}{EA580C}
\newtcolorbox{takeawaybox}{
  colback=takeawaybg,
  colframe=takeawayborder,
  boxrule=0.6pt,
  arc=2.8mm,
  left=2mm,
  right=1.8mm,
  top=1.3mm,
  bottom=1.3mm,
  enhanced,
  breakable,
  borderline west={2.5pt}{0pt}{takeawayaccent}
}

\newcommand{\mypara}[1]{\noindent{\bf {#1}}\xspace.}

\newcommand{\pquote}[2]{\textit{``#1''} (#2)}

\begin{document}

\title{``What Did It Actually Do?'':
Understanding Risk Awareness and Traceability for Computer-Use Agents}

\author{Zifan Peng}
\authornote{Work done during a visit to Newcastle University, The United Kingdom.}
\email{zpengao@connect.hkust-gz.edu.cn}
\orcid{0009-0009-1127-5484}
\affiliation{
  \institution{The Hong Kong University of Science and Technology (Guangzhou)}
  \country{China}
}

\author{Mingchen Li}
\email{MingchenLi@my.unt.edu}
\affiliation{
  \institution{University of North Texas}
  \country{The United States}
}

\renewcommand{\shortauthors}{Peng et al.}

\begin{abstract}
Personalized computer-use agents are rapidly moving from expert communities into mainstream use.
Unlike conventional chatbots, these systems can install skills, invoke tools, access private resources, and modify local environments on users' behalf.
Yet users often do not know what authority they have delegated, what the agent actually did during task execution, or whether the system has been safely removed afterward.

We investigate this gap as a combined problem of risk understanding and post-hoc auditability, using OpenClaw as a motivating case.
We first build a multi-source corpus of the OpenClaw ecosystem, including incidents, advisories, malicious-skill reports, news coverage, tutorials, and social-media narratives.
We then conduct an interview study to examine how users and practitioners understand skills, autonomy, privilege, persistence, and uninstallation.
Our findings suggest that participants often recognized these systems as risky in the abstract, but lacked concrete mental models of what skills can do, what resources agents can access, and what changes may remain after execution or removal. 
Motivated by these findings, we propose \textsc{AgentTrace}, a traceability framework and prototype interface for visualizing agent actions, touched resources, permission history, provenance, and persistent side effects.
A scenario-based evaluation suggests that traceability-oriented interfaces can improve understanding of agent behavior, support anomaly detection, and foster more calibrated trust.
\end{abstract}


\maketitle

\section{Introduction}

Personalized computer-use agents are increasingly presented as practical assistants that can help users install software, run code, browse the web, manage files, and automate multi-step tasks.
Compared with conventional chatbots, these systems are not confined to text generation: they can invoke tools, access private resources, retain state across sessions, and make changes to local or connected environments.
As a result, their central usability challenge is closely tied to security and privacy: users must not only decide whether to trust the system, but also understand what they have delegated and what the agent actually did.

This challenge is becoming more urgent as such systems spread beyond expert communities.
OpenClaw is a useful motivating case because it makes many core properties of personalized computer-use agents unusually visible: skills, tool use, persistent state, and an active surrounding ecosystem of tutorials, deployments, and public discussion~\cite{openclaw_2026,businessinsider2026installation,reuters2026wechat}.
At the same time, recent work shows that risks in these systems are not limited to harmful text outputs, but can emerge through tool invocation, external content, memory, and broad execution authority~\cite{pasb_2026,clawgrip_2026,unsafesearch}.
For end users, however, these risks often appear only as vague unease: they may suspect that the system is dangerous, yet still not know what a skill can execute, whether the agent can act autonomously, or what remains after apparent uninstallation.

This makes personalized computer-use agents an important HCI problem.
Recent studies suggest that users often form simplified mental models of generative-AI ecosystems, especially around third-party extensions and agentic behavior~\cite{wang2025mentalmodels,brachman2025appropriate}.
Meanwhile, visualization and diagnosis research suggests that stepwise and layered traces can improve how people inspect and interpret complex AI workflows~\cite{xie2024waitgpt,lu2025agentlens,dills_2026}.
However, an important gap remains between these two lines of work.
We still know little about how users understand \emph{real} computer-use agent ecosystems, including skills, tutorials, paid installation services, and uninstall concerns; nor do we yet have end-user-centered interfaces for auditing what such agents actually changed.

In this paper, we address this gap by studying personalized computer-use agents as a combined problem of \emph{risk awareness}, \emph{mental models}, and \emph{traceability}.
We use OpenClaw as a motivating case and proceed in three steps.
First, we construct a multi-source corpus of the OpenClaw ecosystem, including incidents, advisories, malicious-skill reports, news stories, tutorials, and social-media narratives, and use this corpus to derive an initial lifecycle-oriented risk taxonomy and ecosystem map.
Second, we conduct an interview study with non-technical users, technical users, and expert deployers to understand how people reason about skills, autonomy, privilege, persistence, and uninstall confidence.
Third, informed by these findings, we propose \textsc{AgentTrace}, a traceability framework and prototype interface that visualizes task timelines, touched resources, permission history, action provenance, and persistent side effects.

Our study points to three recurring tensions.
First, adoption is often shaped by urgency narratives---such as fear of falling behind, pressure to learn AI quickly, or reliance on friends, tutorials, or paid installers---rather than by a clear understanding of the system's authority model.
Second, participants across technical backgrounds could often name ``security'' or ``privacy'' as concerns, yet struggled to explain what a skill could do, what resources an agent could access, or what state might remain after execution or removal.
Third, participants consistently wanted support for post-hoc auditing: rather than only receiving warnings beforehand, they wanted to know what the agent touched, changed, downloaded, opened, and why. 

These findings motivate our central design argument: for personalized computer-use agents, transparency should not stop at prompts, permissions, or chat summaries.
Users also need usable support for reconstructing actions, authority, provenance, and persistence after a task has completed.
To this end, we derive a traceability framework and instantiate it in \textsc{AgentTrace}.
Our evaluation suggests that traceability-oriented interfaces can help participants reconstruct agent behavior, identify risky operations, and plan possible remediation, while also supporting more calibrated trust in agentic systems. 

This paper makes the following contributions:

\begin{itemize}
    \item We contribute a multi-source empirical characterization of the OpenClaw ecosystem as a real-world setting for studying personalized computer-use agents, spanning incidents, malicious skills, tutorials, social narratives, and ecosystem diffusion.
    
    \item We provide qualitative evidence about how different user groups understand---and misunderstand---skills, agent autonomy, system privileges, persistence, and uninstallation in personalized computer-use agents.
    
    \item We propose a traceability framework for personalized computer-use agents that emphasizes five dimensions of post-hoc understanding: task timeline, resource touchpoints, permission history, action provenance, and persistent side effects.
    
    \item We implement and evaluate \textsc{AgentTrace}, a prototype interface showing how behavior traces can improve understanding of agent actions, support anomaly detection, and foster more calibrated trust. 
\end{itemize}

\begin{figure}[htbp]
\centering
\centering
\includegraphics[width=\linewidth]{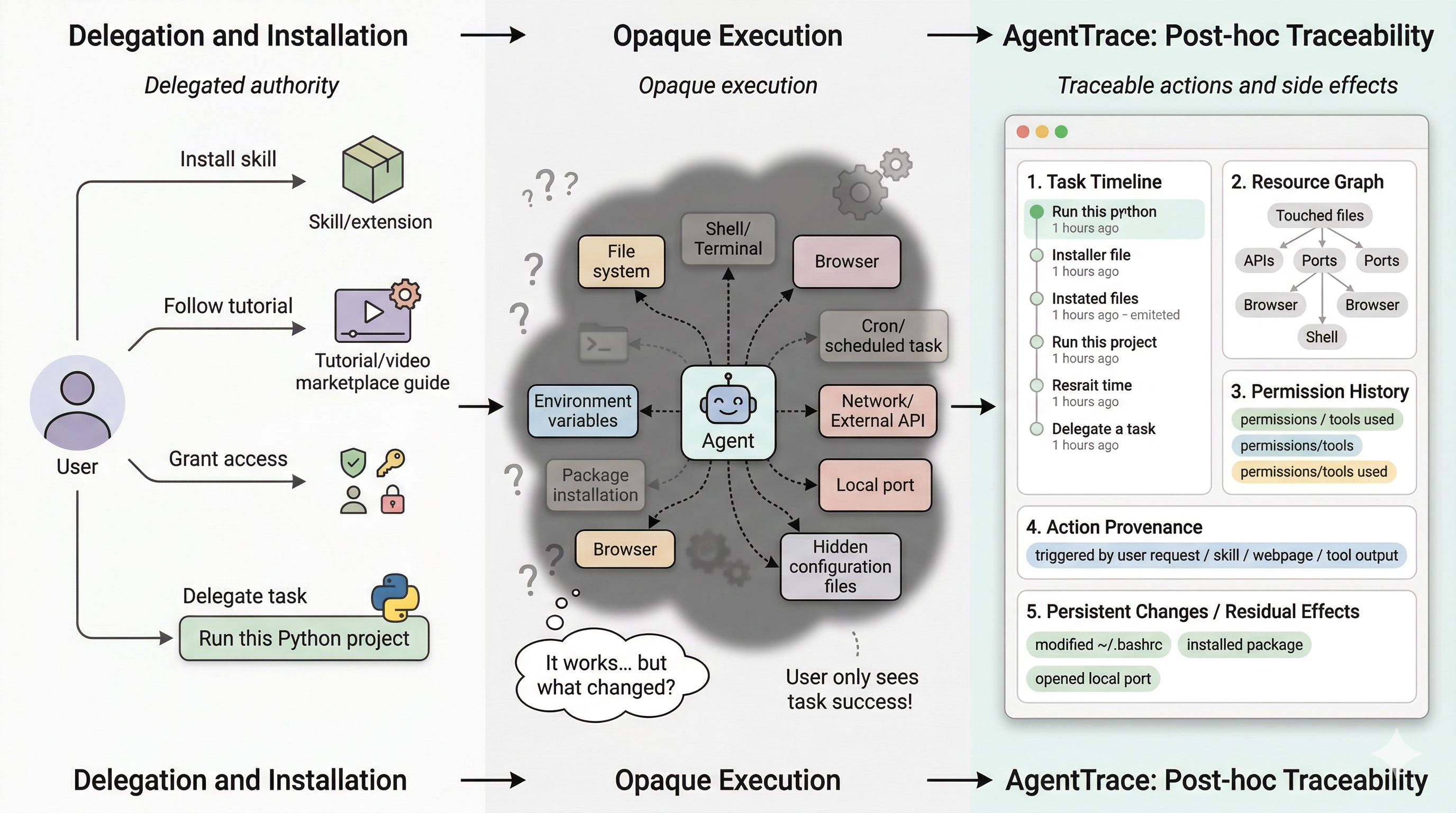}
\caption{Problem framing of this paper. Users delegate tasks and authority to personalized computer-use agents through skills, tutorials, and setup choices, yet the agent’s execution can remain opaque across files, tools, network access, and persistent system changes. We propose AgentTrace, a traceability-oriented interface that makes actions, touched resources, permissions, provenance, and residual side effects legible after task execution.}
\label{fig:teaser}
\end{figure}

More broadly, we argue that the usability problem of personalized computer-use agents is now inseparable from their security and privacy problems.
As these systems become more capable and more socially diffused, the gap between what users think they delegated and what the agent actually did becomes a central HCI challenge.
Closing that gap requires not only better warnings but also better tools for seeing, interpreting, and auditing agent behavior after the fact.
\Cref{fig:teaser} illustrates the problem framing of this paper, from installation and delegated authority to opaque execution and post-hoc traceability needs.

\section{Background and Related Work}

\subsection{Personalized Computer-Use Agents in the Wild}

Personalized computer-use agents differ from conventional chatbots in that they do not merely generate text, but can also invoke tools, access connected resources, retain state across sessions, and act on users' behalf in software environments.
In practice, the relevant unit of interaction is therefore not only a prompt--response exchange, but a broader action loop involving files, shells, browsers, communication channels, credentials, and persistent state.
Recent benchmark work on repository-scale coding agents likewise suggests that the relevant unit of analysis is often a long-horizon action trajectory over a software environment rather than a single-turn prompt--response exchange~\cite{nlp2repo}.
OpenClaw provides a useful motivating case because it makes this structure unusually visible through public documentation, skills, tool invocation, and an active surrounding ecosystem of tutorials, deployments, and public discussion~\cite{openclaw_2026,businessinsider2026installation,reuters2026wechat}.

Recent security work already shows why such systems require dedicated study.
PASB argues that personalized agents should be evaluated with realistic private assets, toolchains, and long-horizon interactions, because failures can propagate across prompt processing, external content, tool invocation, and memory-related behavior~\cite{pasb_2026}.
Similarly, \citet{clawgrip_2026} show that OpenClaw-style local and computer-use agents remain vulnerable across multiple attack categories even when layered safeguards are present.
These works establish personalized computer-use agents as a distinct risk setting rather than simply a more capable chatbot.
At the same time, important evidence about this ecosystem also comes from documentation, advisories, and security reporting on skills, malicious bundles, and deployment practice~\cite{openclawClawhubDocs,meller2026magic,snyk2026toxicskills,trendmicro2026amos}.
In this paper, we use OpenClaw not as the only agent architecture that matters, but as a realistic and well-documented entry point for studying how users understand, trust, and audit high-authority agents in practice.

\mypara{A systems view of personalized computer-use agents}
For the purposes of this paper, we use \emph{personalized computer-use agents} to refer to agentic systems that can continuously interact with users, invoke external tools, access software environments, and retain state across tasks or sessions.
Three system properties are especially important.
First, these agents expose \emph{execution surfaces}, such as filesystem access, command execution, browser interaction, messaging, or device control.
Second, they maintain \emph{persistent state}, including transcripts, memory files, cached artifacts, or automation state, which allows influence to carry across time.
Third, many of them support \emph{extensibility} through skills, plugins, or third-party capability bundles.
This decomposition is useful not only for understanding security risk, but also for explaining why users may struggle to form accurate mental models of what the agent can access, what it changed, and what remains after a task finishes.

\begin{figure}[htpb]
\centering
\includegraphics[width=\linewidth]{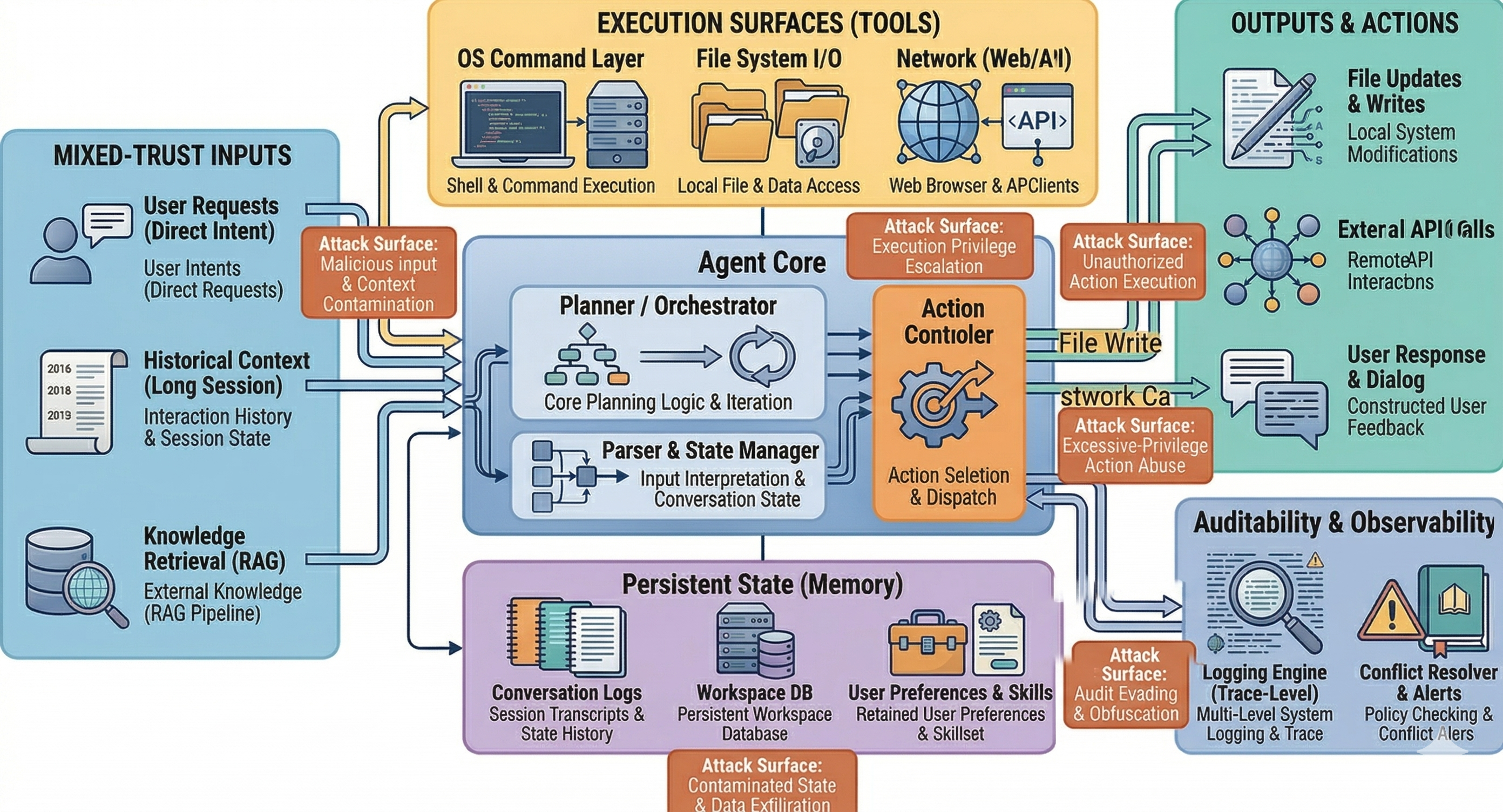}
\caption{Conceptual decomposition of a personalized computer-use agent.
Such systems combine mixed-trust inputs, an agent core, execution surfaces, persistent state, extensibility mechanisms, and user-visible outputs.
This structure helps explain why users may struggle to understand what the agent can access, what it changed, and what remains after task execution.}
\label{fig:prelim_components}
\end{figure}

\subsection{Mental Models, Trust, and Agentic Systems}

A central idea in HCI and usable security is that people act on the basis of their mental models of how a system works.
This is especially important for agentic systems, where users must reason not only about what a system \emph{knows}, but also about what it can \emph{do}, what it can \emph{access}, and how far its actions may continue once delegated.

Recent work begins to examine this problem for generative-AI ecosystems.
\citet{wang2025mentalmodels} study how users understand generative-AI chatbot ecosystems and show that participants often form simplified and internally consistent models of third-party ecosystems, which can in turn be associated with higher trust and fewer concerns.
This result is particularly relevant for systems such as OpenClaw, where third-party skills, tutorials, and deployment pathways are central to adoption.

Relatedly, \citet{brachman2025appropriate} examine what users know and want to know about an agentic AI chatbot.
Their work highlights that users do not automatically hold appropriate mental models of agentic behavior and that they want more actionable information about how the system works, what information it uses, and how it reaches its outputs.
Together, these studies suggest that the challenge of agent adoption is not merely one of interface simplicity; it is also one of helping users form sufficiently accurate mental models of capability, authority, and evidence.

However, prior work in this area has largely focused on chatbot ecosystems or agentic assistants in the abstract.
It remains less clear how users reason about \emph{computer-use agents in the wild}, where delegation is mediated by skills, installation guides, cloud templates, paid setup services, local execution, and persistent system changes.
Our paper extends this line of work by examining risk understanding in a concrete ecosystem where authority is not only conversational, but operational.

\subsection{Transparency, Verification, and Traceability for AI Workflows}

A second relevant line of work concerns how people inspect, verify, and debug AI-supported workflows.
In a study of AI-assisted data analysis, \citet{gu2024verify} show that people often begin verification by reconstructing what the AI actually did before deciding whether the result is correct.
This insight is closely aligned with the problem we study here: when an agent acts through tools and modifies a computing environment, users often need procedural understanding before they can make safety or trust judgments.

Several recent systems make this process more visible.
\citet{xie2024waitgpt} introduce WaitGPT, which transforms LLM-generated code into an on-the-fly, stepwise visual representation so that users can monitor, verify, and steer data-analysis operations.
\citet{lu2025agentlens} propose AgentLens, a visualization system that organizes autonomous-agent event streams into hierarchical temporal summaries and cause traces.
More recently, DiLLS structures multi-agent execution records into layered summaries of activities, actions, and operations to support failure diagnosis~\cite{dills_2026}.

These systems collectively show that traces, summaries, and provenance views can make complex AI behavior more understandable.
Yet they primarily target developer diagnosis, data-analysis verification, or autonomous-system inspection from an expert perspective.
Related efforts in domain-specific settings also suggest the value of decomposing opaque multi-step machine behavior into structured, inspectable explanations for users~\cite{peng2026txsumusercenteredethereumtransaction}.
They do not directly address a different but increasingly important problem: how ordinary or moderately technical users can audit a high-authority computer-use agent after it has changed files, installed dependencies, opened connections, or left persistent state behind.
Our work builds on this literature but reorients it toward \emph{post-hoc auditability for end users and advanced adopters of personalized computer-use agents}.

\subsection{Positioning of This Paper}

This paper sits at the intersection of three strands of work, but is not reducible to any one of them.
First, unlike security benchmark or attack papers on personalized agents~\cite{pasb_2026,clawgrip_2026}, we focus on how people understand real agent ecosystems and what kinds of audit support they need after delegation.
Second, unlike recent mental-model studies of chatbot ecosystems and agentic assistants~\cite{wang2025mentalmodels,brachman2025appropriate}, we study a concrete computer-use setting where skills, persistent state, system modification, and uninstall confidence become central.
Third, unlike trace-visualization and diagnosis systems~\cite{gu2024verify,xie2024waitgpt,lu2025agentlens,dills_2026}, we target user-facing auditability in a high-authority agent context rather than developer debugging alone.

Methodologically, our paper also differs from prior work by combining a multi-source ecosystem corpus, an interview study on risk understanding and uninstall confidence, and a traceability-oriented interface prototype.
This combination allows us to connect three levels that are often studied separately: the public ecosystem through which users encounter computer-use agents, the mental models through which they interpret risk, and the interface mechanisms that may help them reconstruct and govern agent behavior afterward.

\section{Study 0: Corpus Construction and Ecosystem Scaffolding}

Before conducting the interview study, we built a multi-source corpus of the OpenClaw ecosystem in order to characterize how personalized computer-use agents are encountered, discussed, attacked, governed, and normalized in practice.
Our goal was not to produce another broad survey, but to construct a grounded empirical scaffold for the remainder of the paper: a real-world event database, a lifecycle-oriented risk taxonomy, an ecosystem map, and a social-narrative coding set.
These materials served two purposes.
First, they helped us identify concrete scenarios, concepts, and tensions for the interview protocol.
Second, they provided an external empirical frame for interpreting what participants did and did not understand about skills, autonomy, privileges, persistence, and uninstallation.

\subsection{Data Sources and Collection Scope}

We collected materials from five source types: (1) official OpenClaw documentation and platform materials; (2) public vulnerability and advisory sources such as security advisories and incident writeups; (3) security research and threat reports on skills, plugins, or supply-chain abuse; (4) news coverage and ecosystem reporting; and (5) tutorials, discussion posts, and public-facing explanatory materials.
We focused on sources that documented either how OpenClaw and related agents technically operate or how they are socially encountered by users.

The resulting corpus was intentionally heterogeneous.
This was necessary because the OpenClaw ecosystem is evolving faster than peer-reviewed literature alone can capture.
For example, official documentation describes the structure of skills and marketplace moderation; public reporting documents installation events, platform integrations, and the emergence of paid installation and removal services; and security reporting documents malicious skills, supply-chain abuse, and platform responses~\cite{openclaw_2026,openclawClawhubDocs,snyk2026toxicskills,koi2026clawhavoc,trendmicro2026amos,businessinsider2026installation,reuters2026wechat}. 

To improve rigor, we recorded for each source its type, publication date, publisher, relevance to lifecycle stage, and evidence strength.
We treated official documentation, advisories, and firsthand technical reports as stronger evidence for system properties and incidents, while using news coverage, tutorials, and discussion materials primarily to characterize diffusion pathways, narratives, and user-facing framings.
When sources overlapped on the same event, we retained multiple records in order to compare how the event was framed across technical, journalistic, and platform-facing contexts.

\subsection{Real-World Event Database}

We first constructed a standardized event database to document concrete ecosystem events and incident reports.
Each entry corresponded to one event or artifact, such as a malicious-skill campaign, a public advisory, a major installation event, a marketplace governance update, or a publicized uninstall-service phenomenon.
The purpose of this database was not only archival.
It also allowed us to systematically compare what kinds of risks are visible to users before use, during installation, during runtime, and after apparent removal.

For each entry, we recorded a common schema with the following fields:
\emph{object} (e.g., skill, deployment pathway, advisory, social service, platform integration),
\emph{assets at stake} (e.g., files, credentials, browser state, local environment, reputation, money),
\emph{lifecycle stage},
\emph{harm type},
\emph{evidence strength},
and \emph{quotable or citable material}.
Additional metadata included source type, date, region or platform context, and whether the item reflected technical evidence, social diffusion, or governance response.
This standardization made it possible to use the same corpus both for later qualitative interpretation and for deriving design-relevant categories such as persistent side effects, capability import, or uninstall uncertainty.

\subsection{Lifecycle-Oriented Risk Taxonomy}

Using the event database and the broader corpus, we derived a lifecycle-oriented risk taxonomy organized around five stages:
\emph{pre-installation awareness}, \emph{installation and onboarding}, \emph{configuration and capability binding}, \emph{in-use execution}, and \emph{post-use persistence and recovery}.
We adopted a lifecycle view because many salient risks in personalized computer-use agents do not appear at a single technical point.
Instead, they accumulate as authority is discovered, imported, configured, exercised, and only later questioned.

For each stage, we recorded the main attack or confusion surfaces, exposed assets, likely triggering conditions, representative examples, and commonly suggested defensive measures.
For example, installation and onboarding included skills, setup guides, cloud templates, and assisted installation pathways; in-use execution included tool invocation, local environment modification, and externally influenced behavior; and post-use recovery included uninstall confidence, residual dependencies, local configuration changes, and uncertainty about what remained after removal.
This taxonomy later served as a sensitizing framework for the interview study, especially when probing participants about what they believed a skill does, what an agent can keep doing after a task starts, and whether uninstalling the application is enough to make the system ``gone.''

\subsection{Lifecycle-Oriented Risk Taxonomy}

To organize the corpus in a way that is useful for both qualitative inquiry and interface design, we developed a lifecycle-oriented risk taxonomy for personalized computer-use agents.
Rather than treating failures as isolated prompt errors, this taxonomy frames risk as a progression of delegated authority: capabilities are discovered before use, imported during installation, bound to concrete assets during configuration, exercised through tools at runtime, and only partially reconstructed after the fact.
This framing was especially useful for our study because participants often struggled not only with what an agent could do in principle, but also with when authority was granted, how it expanded, and what remained afterward.

We organized the taxonomy around five stages:
\emph{pre-installation awareness},
\emph{installation and onboarding},
\emph{configuration and capability binding},
\emph{in-use execution},
and
\emph{post-use persistence and recovery}.
For each stage, we recorded the main attack or confusion surfaces, exposed assets, triggering conditions, representative examples, and typical protective measures.
This lifecycle lens allowed us to connect technical risk with user-facing experience.
For example, installation is where users first import capabilities they may not fully understand, configuration is where abstract capability becomes concrete authority, runtime is where mixed-trust content meets execution, and post-use is where users must determine what changed and what still remains.

\Cref{tab:lifecycle_taxonomy} summarizes this lifecycle-oriented taxonomy.
We used this taxonomy as a sensitizing framework for the interview protocol, especially when probing how participants understood skills, autonomy, privileges, persistence, and uninstall confidence.
It also later informed the design of \textsc{AgentTrace}, particularly its focus on touched resources, authority history, provenance, and persistent side effects.

\begin{table}[htbp]
\centering
\caption{Lifecycle-oriented risk taxonomy for personalized computer-use agents. The taxonomy connects ecosystem evidence, user-facing confusion points, and system-level consequences across five stages of delegated authority.}
\label{tab:lifecycle_taxonomy}
\small
\begin{tabularx}{\textwidth}{p{2cm} p{2.5cm} p{2.5cm} p{2.8cm} p{3.5cm} X}
\toprule
\textbf{Stage} & \textbf{Main Surface} & \textbf{Exposed Assets} & \textbf{Triggering Conditions} & \textbf{User-Facing Consequences} \\
\midrule
Pre-installation awareness
& News, tutorials, peer recommendation, paid setup services
& Trust, decision quality, account access delegated to helpers
& Urgency narratives, fear of missing out, low technical confidence
& Adoption before clear understanding of what the system can access or modify \\

Installation and onboarding
& Skills, plugins, dependency setup, setup guides
& Files, credentials, future execution surface
& One-command install, third-party skills, natural-language onboarding
& Users import capability bundles without clearly understanding whether they execute code, install dependencies, or expand authority \\
\midrule

Configuration and capability binding
& Filesystem scope, browser profiles, channels, keys, nodes, sandbox settings
& Concrete local and connected resources
& Broad defaults, convenience setup, weak isolation, channel binding
& Users cannot easily judge what the agent is actually authorized to touch in a given deployment \\

In-use execution
& Tool invocation, mixed-trust content, autonomous planning
& Files, browser state, services, messages, local environment
& Local task execution, external content, skill-defined steps, multi-step continuation
& Users see task progress but may not notice side effects, risky operations, or escalation of scope \\
\midrule

Post-use persistence and recovery
& Logs, memory, installed dependencies, config changes, residual services
& Persistent state, privacy, environment integrity
& Task completion, uninstall attempts, later troubleshooting
& Users are unsure what changed, what remains, and whether the system has been fully removed or requires remediation \\
\bottomrule
\end{tabularx}
\end{table}

\subsection{Ecosystem Map and Social-Narrative Coding}

We also used the corpus to build an OpenClaw ecosystem map.
Rather than treating OpenClaw as a single software artifact, the map captured relationships among the official agent runtime, platformized variants and integrations, skill marketplaces, tutorials, paid setup services, technically experienced helpers, and ordinary users.
This was important because participants often encounter agent systems not through source code or official security documentation, but through social pathways such as installation events, recommendation chains, cloud deployment guides, and third-party explainers~\cite{businessinsider2026installation,reuters2026wechat}. 

In parallel, we conducted a lightweight content analysis over news stories, tutorials, and public-facing discussions in order to identify recurring narrative themes.
Our initial coding tracked framings such as productivity gain, fear of falling behind, entrepreneurship or side-hustle narratives, low-friction installation, security panic, uninstall anxiety, and organizational restriction.
These themes were not treated as the paper's main contribution by themselves.
Instead, they helped us understand the interpretive environment in which participants encountered computer-use agents and the kinds of assumptions they may already have formed before using or discussing such systems.

\subsection{Role of Study 0 in the Remainder of the Paper}

Study 0 functioned as a formative empirical scaffold for the rest of the project.
It informed the interview protocol in three ways.
First, it helped us construct realistic prompts and vignettes, such as third-party skill installation and local project execution with potentially persistent side effects.
Second, it informed the initial sensitizing concepts for coding interview data, including capability import, assisted installation, selective protection, uninstall confidence, and post-hoc audit needs.
Third, it highlighted which aspects of agent behavior users might plausibly need to reconstruct after execution, such as execution order, modified resources, privilege context, and persistent residual changes.

In this sense, Study 0 did not stand apart from the later user study and prototype design.
It provided the ecological grounding that allowed us to move from public ecosystem evidence to user mental models, and from user mental models to the design of traceability-oriented interfaces.

\section{Study 1: Interview Study on Risk Awareness and Audit Needs}

To understand how people perceive personalized computer-use agents in practice, we conducted an interview study focused on risk awareness, trust, and post-hoc audit needs.
Our goal was not only to identify what participants found concerning, but also to understand how they reasoned about skills, autonomy, system privileges, persistence, and uninstall confidence in the context of OpenClaw and similar systems.
Building on the ecosystem scaffold introduced in the previous section, this study examined how public narratives, installation pathways, and prior technical experience shaped participants' mental models of what these systems can do and how they should be trusted or constrained.

\subsection{Setup}

\mypara{Recruitment and screening}
We recruited participants through university mailing lists, social media posts, online communities related to AI tools, and personal referrals.
Our goal was to cover a range of technical backgrounds and usage pathways, including participants who had directly used OpenClaw or similar computer-use agents, participants who had seriously considered using them, and expert users who had deployed, inspected, or discussed such systems in practice.
During screening, we asked about participants' familiarity with LLM tools, whether they had used AI agents beyond chatbots, whether they had installed or configured skills, plugins, or local deployment environments, and whether they had concerns about what such systems could access or modify.

\mypara{Participants}
We interviewed 16 participants in total (P1--P16), aged from 20 to 38.
To capture a range of perspectives, we recruited participants from three broad groups: non-technical or low-technical users, technical users with hands-on experience using or configuring AI tools, and expert deployers such as developers, security-aware users, or practitioners who had inspected or deployed agent systems in real settings.

Several participants had firsthand experience using OpenClaw or closely related systems, while others had encountered them through tutorials, news reports, installation services, or discussions with technically experienced peers.
Across groups, participants varied in how often they used LLMs, how much they understood local environments and permissions, and whether they had previously relied on others to help them install or configure tools.

\mypara{Procedure}
We conducted semi-structured interviews covering five topics: first exposure to computer-use agents, understanding of system capabilities and risks, trust decisions about skills or deployment pathways, beliefs about what an agent may have changed after execution, and desired forms of audit support.
We asked participants how they first encountered OpenClaw or similar systems, whether they felt pressure or urgency to adopt them, and how they imagined the installation process.
We then probed how they understood skills, autonomy, external influence, and access to concrete assets such as files, browser state, API keys, shell configuration, environment variables, and background tasks.

To ground the discussion, we used two scenario prompts: a third-party skill installation scenario and a local Python-project execution scenario involving possible dependency installation, configuration changes, and persistent side effects.
We also showed a small set of low-fidelity audit mockups, including a task timeline, modified-resource list, permission history, and persistent-change summary, and asked participants which views were most useful for determining what happened and whether remediation was necessary.
This final step helped us move from abstract risk perception to concrete design implications for traceability-oriented interfaces.

\mypara{Analysis}
All interviews were recorded and transcribed into text.
We analyzed the transcripts using reflexive thematic analysis.
We first coded participants' exposure pathways, mental models, trust cues, protective practices, failure points in understanding, and desired forms of audit support.
We then grouped these codes into broader themes related to risk awareness, skill understanding, privilege understanding, uninstall confidence, and post-hoc traceability needs.
Throughout the analysis, we compared patterns across participant groups to identify both shared concerns and differences between non-technical users, technical users, and expert deployers.

\subsection{Findings}

We organize the findings around four questions from the interviews, covering why participants turned to these systems, how they understood their risks and capabilities, how they decided whether to trust them, and why they wanted stronger support for auditing agent behavior.

\subsubsection{Why do users turn to personalized computer-use agents?}
Participants rarely described adoption as a purely curiosity-driven choice. Instead, many framed it as a response to urgency, pressure, or fear of falling behind.

\begin{takeawaybox}
\textbf{Finding 1.} Participants (14/16) often approached computer-use agents through urgency, pressure, or social influence rather than through deliberate security evaluation.
\end{takeawaybox}

Participants rarely described adoption as a purely curiosity-driven choice.
Instead, many framed it as a response to pressure: pressure to keep up with rapid AI adoption, to remain employable, or to avoid being left behind by colleagues or the broader technological environment.
For these participants, the decision to try OpenClaw or similar systems began from urgency rather than from a careful comparison of risk and benefit.
\pquote{I did not start from asking whether it was safe. I started from feeling that if I did not learn it now, I would be left behind.}{P5} 
Some participants framed this urgency in explicitly competitive or workplace terms.
They described AI agents as something they felt they were expected to understand, even before they had formed a clear idea of what the system could actually access or modify.
\pquote{It was more like, everyone around me was talking about these agents as the next thing. I felt I at least needed to know how they worked, otherwise I would become outdated.}{P12} 
At the same time, not all participants described urgency in the same way.
For a smaller set of participants, the attraction was not fear but novelty and experimentation.
These participants were drawn in because the systems seemed powerful, playful, or simply interesting to try.
\pquote{For me it was not fear first. It was more like, this sounds crazy, I want to see whether it can actually do things on my computer.}{P3} 
Together, these accounts suggest that adoption was often shaped by social narratives---such as AI pressure, competitive anxiety, or curiosity amplified by online attention---rather than by a stable understanding of the system's authority model.

\begin{takeawaybox}
\textbf{Finding 2.} Low-friction (2/16) and assisted installation lowered the threshold for adoption, but also displaced trust toward other people and platforms.
\end{takeawaybox}

Participants described a wide range of entry pathways, including self-installation, following online tutorials, using cloud deployment guides, asking technically experienced friends for help, or paying others to install and configure the system.
This pattern was especially visible among non-technical users, who often treated installation as a service or a procedural hurdle to get past, rather than as a moment of security-relevant decision making.
\pquote{I would probably find someone to help me install it first, because I would not know what those commands or settings really mean.}{P2} 
In these cases, trust was often transferred away from the system itself and onto intermediaries such as tutorial authors, friends, deployment platforms, or installers.
Participants sometimes admitted that they would trust a system mainly because someone more technical had said it was fine, not because they themselves understood what was happening.
\pquote{If a friend who knows these things helps me set it up, I would probably trust that more than trying to inspect it myself.}{P6} 
This suggests that installation is not only a technical step but also a social one: low-friction onboarding can increase adoption while simultaneously reducing users' direct engagement with the system's risks and boundaries.

\subsubsection{How do users understand skills, autonomy, and agent capabilities?}
Although many participants recognized that such systems might be risky, their understanding of how those risks arise was often incomplete.

\begin{takeawaybox}
\textbf{Finding 3.} Participants (12/16) often recognized abstract risk, but lacked concrete mental models of what a skill can do.
\end{takeawaybox}

Across groups, participants frequently described skills as potentially dangerous, but many could not clearly explain whether skills were closer to plugins, prompts, scripts, or software packages.
Some imagined them as reusable instructions or convenience templates.
Others suspected that they might involve code execution, but could not specify what kinds of code, dependencies, or privileges might be involved.
\pquote{I knew a skill might be risky, but I was not sure whether it was actually running code or just telling the agent what to do.}{P8} 
For some participants, the uncertainty was not whether skills were safe, but what category of thing they even belonged to.
This ambiguity made it difficult for them to reason about installation risk in any concrete way.
\pquote{When you say ‘skill’, I do not know whether I should imagine an extension, a script, or just a smarter prompt. Those feel like very different levels of danger.}{P10} 
Even among technically experienced users, direct inspection was often selective and partial.
Participants might know that a skill could be risky, but still rely on surface impressions or prior assumptions rather than a full understanding of capability import.

\begin{takeawaybox}
\textbf{Finding 4.} Users underestimated how much autonomy and authority an agent may have once a task begins.
\end{takeawaybox}

Many participants initially described the agent as an advanced chatbot.
However, this framing became unstable as we probed specific actions such as editing files, changing environment variables, opening ports, installing packages, or continuing across multiple steps without explicit confirmation.
Participants often revised their understanding during the interview, realizing that the system's practical authority might be broader than they had assumed.
\pquote{At first I thought it was mostly about files and commands, but when you mentioned things like \texttt{.bashrc}, ports, or browser passwords, I realized I did not really know where the boundary was.}{P11} 
Several participants could name one or two sensitive assets, such as files or API keys, but overlooked others such as browser state, shell configuration, scheduled tasks, or residual local services.
This suggests that their mental models of privilege were fragmented rather than absent.
\pquote{I would have thought about documents and API keys, but not about the system changing environment variables or leaving something running in the background.}{P4} 
Participants also varied in how much autonomy they attributed to the agent.
Some assumed that the system would ask before each meaningful action.
Others only realized during the scenarios that the agent might continue through multiple steps once the task had started.
\pquote{I think I was assuming it would kind of stop and ask me a lot. But if it keeps going after the first instruction, that feels very different.}{P7} 

\subsubsection{How do users decide whether to trust a skill or deployment pathway?}
Participants did not rely on one stable trust criterion. Instead, they used a mix of heuristics, reputational cues, and practical shortcuts.

\begin{takeawaybox}
\textbf{Finding 5.} Trust decisions were often based on ecosystem cues rather than direct inspection.
\end{takeawaybox}

Participants did not report using a single stable trust criterion.
Instead, they relied on a mixture of popularity, apparent professionalism, recommendations, source appearance, and whether something felt ``official.''
Few participants said they would regularly inspect skill code or dependency specifications in detail, and even technically experienced users described such inspection as selective rather than routine.
\pquote{In practice, I would probably trust the source more than inspect every line, unless something already looked suspicious.}{P13} 
Some participants explicitly said they would look for signs such as how widely a tool seemed used, whether a tutorial was polished, or whether a marketplace looked established.
\pquote{If it looks like a popular skill and the tutorial is well made, I would probably assume it is reasonably safe unless I saw some obvious red flag.}{P14} 
This means that trust was frequently inferred from surrounding ecosystem signals rather than from direct understanding of what the agent or skill would actually do.

\begin{takeawaybox}
\textbf{Finding 6.} Protective practices were selective, uneven, and often only partially understood.
\end{takeawaybox}

Some technical participants mentioned Docker, sandboxing, account separation, or dedicated environments as protective strategies.
However, even when participants could name such measures, they did not always understand what those protections covered or what they failed to cover.
For others, protection was much more ad hoc: trusting the installer, using a secondary account, or hoping that uninstalling the tool would be enough.
\pquote{I know people say to use Docker or a sandbox, but honestly I could not explain exactly what that would protect me from in this case.}{P9} 
Among non-technical users in particular, protection was often imagined as delegation: if someone more knowledgeable handled the installation, then the risk was assumed to be lower.
\pquote{My protection would probably be asking someone technical to help, because I would not know how to protect myself properly.}{P1} 
These accounts suggest that users often recognized the existence of protective measures without possessing a strong operational understanding of how they mapped onto the actual risks of computer-use agents.

\subsubsection{Why do users want auditability after execution?}
One of the most consistent themes across interviews was that participants wanted to reconstruct what had happened after the task was over.

\begin{takeawaybox}
\textbf{Finding 7.} Participants were often unsure whether they could tell what the agent had changed.
\end{takeawaybox}

When discussing concrete scenarios such as running a Python project, installing a skill, or connecting a service, participants repeatedly said that the hardest part was not only deciding whether to start, but also knowing what had happened afterward.
They worried that the agent might have modified the local configuration, downloaded extra dependencies, written to unknown directories, opened ports, or left behind persistent state without making those changes visible.
\pquote{The scary part is not just whether it did the task. It is that afterward I may have no idea what else it changed.}{P16} 
Some participants focused on environment drift.
They were less worried about one catastrophic action than about many small changes accumulating in ways they would not know how to detect later.
\pquote{What bothers me is not only a big obvious mistake. It is the possibility that it quietly changed several things, and later I cannot even tell why my setup is different.}{P15} 
This pattern appeared across technical backgrounds, although participants differed in how precisely they could name the kinds of changes they were worried about.

\begin{takeawaybox}
\textbf{Finding 8.} Participants lacked confidence in uninstalling or fully removing the system.
\end{takeawaybox}

A recurring concern was that even if the main application was removed, participants could not be sure whether related files, installed dependencies, credentials, environment variables, or background tasks had also been removed.
This concern was especially strong among participants with less technical background, but it also appeared among technical users who were aware that agent systems may modify multiple parts of a local environment.
\pquote{Even if I uninstall it, I would not know whether it is really gone, or whether something it changed is still there.}{P1} 
For some participants, uninstall confidence was tied to hidden residue: things that may still exist but are no longer visible in the main interface.
\pquote{Removing the app is one thing. But how would I know whether it added packages, changed config, or left some credential file somewhere?}{P6} 
This suggests that participants did not equate removal of the visible application with restoration of the previous system state.

\begin{takeawaybox}
\textbf{Finding 9.} Participants wanted post-hoc auditing rather than only pre-action warnings.
\end{takeawaybox}

Across technical backgrounds, participants consistently asked for interfaces that could show what the agent touched, changed, downloaded, opened, and why.
When reacting to our mockups, they especially valued representations of execution order, modified resources, persistent side effects, and the source of each action.
Rather than asking only for more warning popups, they wanted support for reconstructing what happened and deciding whether remediation was needed.
\pquote{I do not just want the system to warn me before it acts. I want to know afterward exactly what it did, and whether I should undo something.}{P2} 
Some participants emphasized sequence and causality: they wanted to know not only what changed, but in what order and because of which input or instruction.
\pquote{If I see that it changed five things, I also want to know what led to each one. Otherwise I still cannot tell what was intentional and what was weird.}{P11} 
Others prioritized persistence and reversibility.
They said that a useful interface should help them identify what remains after the task ends and what should be rolled back first.
\pquote{The most useful thing for me would be a summary of what is still there now, not just a log of what happened before.}{P3} 
Taken together, these reactions suggest that auditability is not a secondary debugging feature, but a central user requirement for personalized computer-use agents.

\section{AgentTrace: A Traceability Framework and Prototype}


The interview study suggests that users need more than pre-action warnings when interacting with personalized computer-use agents.
Participants wanted to understand what a skill imports, what resources an agent can reach once execution begins, what changes remain after task completion, and whether any persistent side effects survive apparent removal.
Across technical backgrounds, they repeatedly expressed a need for support in reconstructing what the agent did, why those actions occurred, and whether follow-up remediation was necessary.
These findings motivated the design of \textsc{AgentTrace}, a traceability framework and prototype interface for post-hoc auditing of personalized computer-use agents.

Rather than treating transparency as a matter of showing prompts or final summaries alone, \textsc{AgentTrace} treats traceability as a first-class design requirement.
The framework is intended to support post-hoc understanding of five aspects of agent behavior: executed actions, touched resources, authority context, triggering context, and persistent residual changes.
This framing builds on prior work showing that users benefit from stepwise views of AI-assisted workflows and layered summaries of agent behavior, but reorients these ideas toward end-user auditability in high-authority agent systems~\cite{xie2024waitgpt,lu2025agentlens,dills_2026,gu2024verify}.

\begin{figure*}[t]
\centering
\includegraphics[width=1\textwidth]{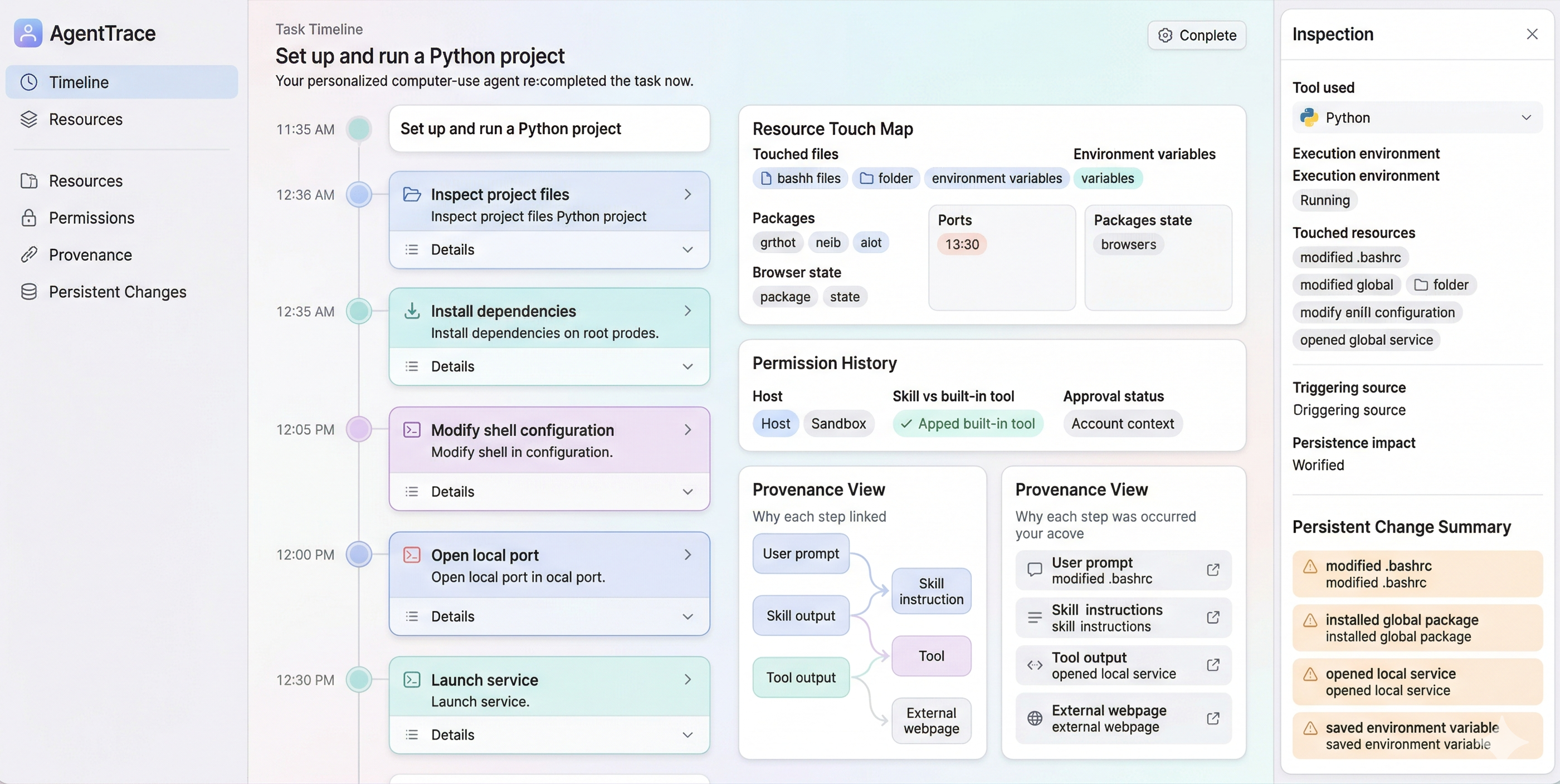}
\caption{\textsc{AgentTrace}, our traceability-oriented prototype for personalized computer-use agents. 
The interface combines five coordinated views for post-hoc auditing: a task timeline, a resource touch map, a permission and authority history, an action provenance inspector, and a persistent change summary. 
Together, these views help users reconstruct what the agent did, what it touched, under what authority it acted, why actions occurred, and what residual changes remained after execution.}
\label{fig:agenttrace_interface}
\end{figure*}

\Cref{fig:agenttrace_interface} shows the overall interface of \textsc{AgentTrace}, which organizes post-hoc audit information into five coordinated views.

\subsection{Overview and Design Rationale}

\textsc{AgentTrace} is designed around five requirements derived from our findings:

\begin{itemize}[leftmargin=*]
    \item \textbf{Legible capability import.} Users should be able to tell when a task depends on third-party skills, added dependencies, or setup steps that expand agent authority.

    \item \textbf{Legible authority boundaries.} The interface should show under what authority each action occurred, including the tool, environment, account, and approval context.

    \item \textbf{Reconstructable execution.} Agent behavior should be presented as a sequence of concrete actions rather than only as a chat-style summary.

    \item \textbf{Visible persistence.} The interface should foreground durable side effects such as modified files, environment changes, installed dependencies, open services, and scheduled tasks.

    \item \textbf{Actionable remediation.} The interface should help users identify what may need review, cleanup, or rollback after task completion.
\end{itemize}

\begin{figure}[htbp]
\centering
\includegraphics[width=\linewidth]{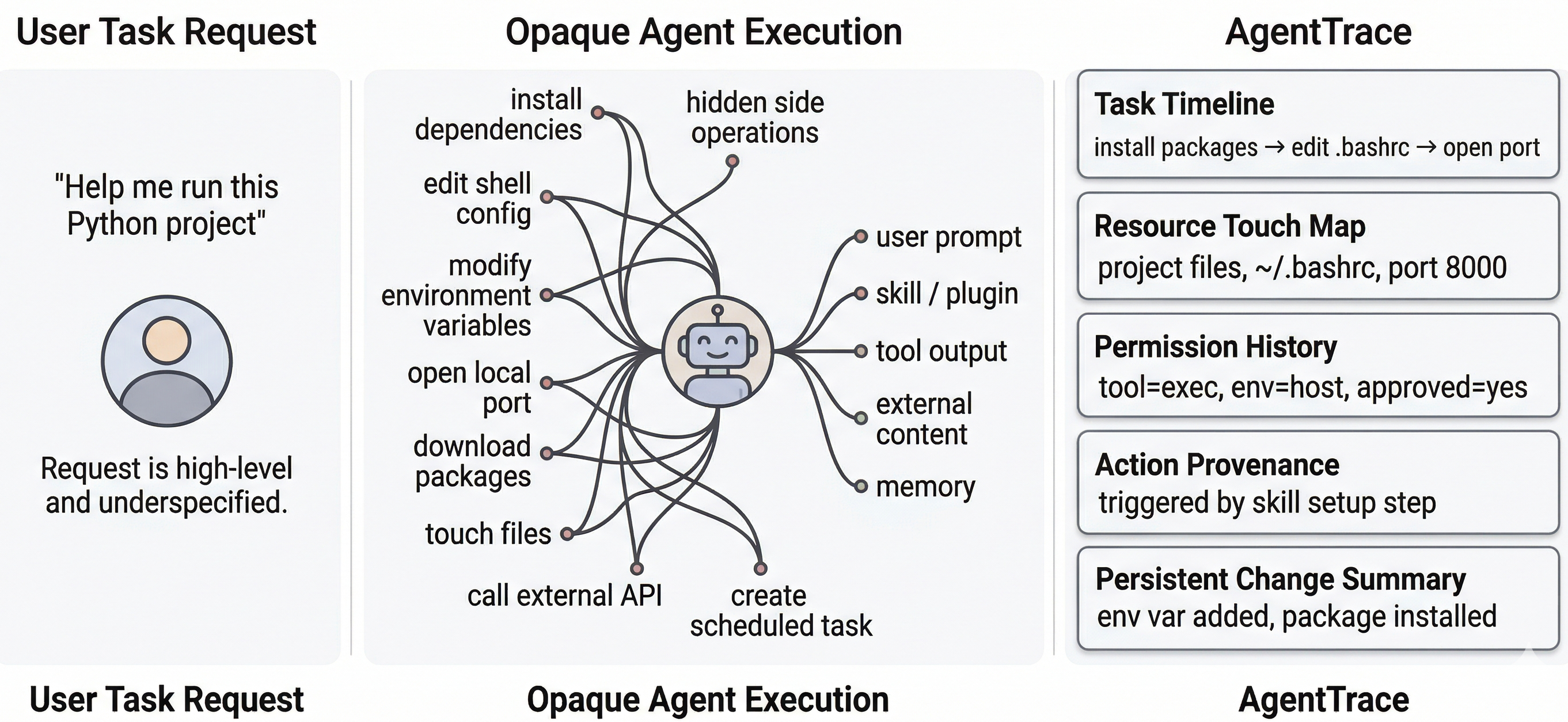}
\caption{\textsc{AgentTrace} turns opaque agent execution into post-hoc audit support. Starting from a high-level user request, personalized computer-use agents may perform multi-step operations involving tools, imported skills, external content, and persistent system changes. \textsc{AgentTrace} organizes this behavior into five coordinated views---task timeline, resource touchpoints, permission history, action provenance, and persistent change summary---to help users reconstruct what happened and determine whether follow-up review or remediation is needed.}
\label{fig:agenttrace_overview}
\end{figure}

Guided by these requirements, \textsc{AgentTrace} organizes post-hoc audit information into five trace dimensions: task timeline, resource touchpoints, permission history, action provenance, and persistent side effects.
Together, these dimensions bridge the gap between a user's high-level task request and the often opaque sequence of system-level operations underneath.
\Cref{fig:agenttrace_overview} illustrates the design rationale of \textsc{AgentTrace}, highlighting how the system transforms opaque multi-step agent execution into user-facing audit information.

\subsection{Trace Model}

To support these five dimensions, \textsc{AgentTrace} represents agent execution as a structured trace composed of five core entities: \emph{actions}, \emph{resources}, \emph{authorities}, \emph{triggers}, and \emph{persistence deltas}.
This model is intended to be lightweight enough to apply across computer-use agents while still expressive enough to surface security- and audit-relevant details.

\mypara{Actions}
An action is a discrete operation taken by the agent, such as reading a file, editing a configuration file, invoking a shell command, downloading a dependency, opening a port, sending a message, or installing a package.
Each action records its timestamp or order, type, status, and relationship to the broader task.

\mypara{Resources}
A resource is any object touched during execution, including files, directories, environment variables, browser state, local services, credentials, packages, domains, ports, or communication targets.
Resources are used to summarize what the agent interacted with and to distinguish between expected and unexpected surfaces of contact.

\mypara{Authorities}
An authority record captures the context under which an action occurred.
This may include the tool used, the execution environment (e.g., host, sandbox, container, remote node), the account or identity involved, whether the action required prior approval, and whether the action occurred under an imported skill or built-in capability.
This dimension addresses participants' recurring uncertainty about where the real privilege boundary lies.

\mypara{Triggers}
A trigger records why an action happened.
Triggers may include the user's original instruction, a skill-provided setup step, a tool return value, an external webpage or document, a memory retrieval, or a follow-on planning step.
This dimension is important because participants wanted to know not only what happened, but why the system believed the action was appropriate.

\mypara{Persistence deltas}
A persistence delta captures any change that remains after the immediate action sequence is complete.
Examples include installed dependencies, modified files, saved configuration, created scheduled tasks, opened services, cached credentials, or other residual state.
By separating persistence from one-off actions, \textsc{AgentTrace} emphasizes the difference between transient execution and durable system modification.

Together, these entities allow the interface to move beyond raw logs.
Instead of presenting an opaque event stream, \textsc{AgentTrace} turns execution into a user-facing representation of what the agent did, what it touched, why it acted, and what remains.

\subsection{Prototype Interface}

We instantiated the framework as a low- to medium-fidelity prototype interface called \textsc{AgentTrace}.
The prototype is organized into five coordinated views, each corresponding to one of the trace dimensions above.

\subsubsection{Task Timeline}

The task timeline presents agent behavior as an ordered sequence of high-level steps.
Each step groups one or more lower-level operations into a human-readable event, such as \emph{inspect project}, \emph{install dependencies}, \emph{modify local configuration}, or \emph{launch service}.
This view is meant to address participants' desire to reconstruct what happened in temporal order rather than reading an undifferentiated log dump.

Each timeline entry includes a concise description, a timestamp or sequence position, a severity or review marker where relevant, and expandable details showing the lower-level actions contained within that step.
Potentially risky actions---such as modifying shell configuration, writing outside the project directory, opening a port, or installing global packages---can be visually emphasized. 

\subsubsection{Resource Touch Map}

The resource touch map summarizes what the agent interacted with during execution.
Rather than listing only commands or outputs, this view aggregates touched files, directories, environment variables, ports, domains, packages, credentials, browser artifacts, and communication targets.
Resources can be grouped by type or sensitivity, allowing users to distinguish expected task-related resources from broader system changes.

This view is intended to help users answer questions such as:
Did the agent stay inside the project directory?
Did it touch browser state?
Did it install anything globally?
Did it open any network-facing surfaces?
By making touched resources explicit, the interface helps users inspect the scope of execution rather than inferring it indirectly from text summaries.

\subsubsection{Permission and Authority History}

The permission history view records the authority context of each action.
For every major step, it shows which tool was used, whether the action occurred on the host or in an isolated environment, whether it involved a third-party skill, whether it was pre-approved or user-confirmed, and which identity or account context was involved.

This view is motivated by a recurring pattern in our interviews: participants often confused what the system was \emph{capable of} in principle with what it was \emph{authorized to do} in a given session.
Permission history therefore makes authority explicit and time-local, helping users see not only that an action occurred, but under what conditions it was allowed to occur.

\subsubsection{Action Provenance Inspector}

The provenance view explains why a step occurred by linking actions to their most relevant upstream trigger.
Depending on the case, this may be the user's prompt, a skill setup instruction, a tool output, a prior plan step, a retrieved memory item, or externally sourced content.
The goal of this view is not to fully solve causality, but to provide enough provenance to support user understanding and suspicion.

For instance, if a package installation followed a skill-defined setup requirement, or if a browser action followed instructions extracted from a webpage, the provenance view can expose that relationship.
This directly addresses a concern repeatedly raised in the interviews: users wanted to know not only what the agent did, but what led it to do that.

\subsubsection{Persistent Change Summary}

The persistent change summary isolates system modifications that are likely to outlast the immediate task.
This includes modified files, added environment variables, installed packages, opened services, saved credentials, created scheduled tasks, and other durable artifacts.
Each item is presented with a brief rationale for why it matters and, where appropriate, a cue that it may require later inspection or rollback.

This view is central to the uninstall and residue concerns voiced by participants.
Many said they could imagine removing the main application yet still not knowing whether related files, dependencies, or services remained.
By foregrounding persistent change, \textsc{AgentTrace} helps users reason about what is still present even after the task is complete.

\subsection{Interface Workflow}

In use, \textsc{AgentTrace} supports a layered workflow.
Users can begin with the high-level task timeline to understand the overall arc of execution.
From there, they can pivot into the resource touch map to inspect scope, the permission history to inspect authority, the provenance view to inspect triggers, and the persistent change summary to inspect residue.
This structure is meant to support both broad situational awareness and targeted investigation.

The prototype therefore supports multiple styles of post-hoc questioning.
A user can ask:
\emph{What happened overall?}
\emph{What did it touch?}
\emph{What looked risky?}
\emph{Why was this step taken?}
or
\emph{What do I need to clean up?}
Rather than forcing one debugging strategy, \textsc{AgentTrace} is designed to accommodate the more exploratory audit behavior that participants described in our interviews.

\subsection{Implementation Sketch}

Our prototype is currently implemented as an interface mockup backed by a structured event schema inspired by real computer-use agent traces. 
At a conceptual level, the system assumes access to three categories of information:
(1) agent action logs or tool events,
(2) resource-level deltas such as file modifications, package installations, or environment changes,
and
(3) contextual metadata such as tool identity, skill involvement, approval state, and source prompt or trigger information.

Although our motivating case is OpenClaw, the framework is intended to be more general.
The trace model does not depend on a specific runtime or marketplace.
Instead, it assumes that agent behavior can be represented as structured actions over resources under specific authority contexts, with some subset of those actions leaving persistent residue.
This makes \textsc{AgentTrace} potentially applicable beyond OpenClaw to other classes of personalized or high-authority computer-use agents. 

\mypara{Summary}
\textsc{AgentTrace} translates our interview findings into a traceability-oriented prototype for personalized computer-use agents.
Where participants expressed uncertainty about skills, authority, and uninstall residue, the system foregrounds capability import, permission context, action provenance, and persistent changes.
Where participants asked for support in determining what happened after the fact, the system provides coordinated views for reconstructing execution at multiple levels of detail.

\section{Evaluation}

We conducted a scenario-based user study to examine whether \textsc{AgentTrace} helps people better understand, inspect, and assess the behavior of personalized computer-use agents after task execution.
Our goal was not to evaluate a full deployment pipeline, but to test whether traceability-oriented views improve users' ability to reconstruct what the agent did, identify risky operations, and judge whether remediation may be needed.

\subsection{Research Questions}

The evaluation addressed four questions:

\begin{itemize}
    \item \textbf{RQ1:} Does \textsc{AgentTrace} improve users' understanding of what the agent did during task execution?
    \item \textbf{RQ2:} Does \textsc{AgentTrace} improve users' ability to identify risky or unexpected actions?
    \item \textbf{RQ3:} Does \textsc{AgentTrace} improve users' ability to determine what may require follow-up inspection or rollback?
    \item \textbf{RQ4:} Does \textsc{AgentTrace} support more calibrated trust and stronger perceived control than baseline summaries or logs?
\end{itemize}

\subsection{Study Design}

We used a within-subject design with two interface conditions:

\begin{itemize}
    \item \textbf{Baseline:} participants viewed a conventional post-task summary consisting of a chat-style execution summary together with a simplified textual log.
    \item \textbf{\textsc{AgentTrace}:} participants viewed the same task through the \textsc{AgentTrace} interface, including the task timeline, resource touch map, permission history, provenance inspector, and persistent change summary.
\end{itemize}

Each participant completed three task scenarios under both conditions.
We counterbalanced condition order and scenario order to reduce sequence effects.
This design allowed us to compare the same participant's performance when inspecting the same type of agent behavior with and without traceability-oriented support.

\subsection{Participants}

We recruited 
12 participants for the evaluation, including both technically experienced and less technical users.

Participants were different from those in the interview study. 
All participants had prior familiarity with LLM-based tools, but their experience with computer-use agents varied.

\subsection{Scenarios and Tasks}

We designed three realistic scenarios based on the corpus analysis and interview findings.

\mypara{Scenario 1: Running a local Python project}
Participants were told that the user had asked an agent to ``get this Python project running.''
The underlying trace included actions such as installing dependencies, editing shell configuration, changing environment variables, and downloading additional packages.
One or more operations were designed to be plausibly useful but potentially risky, such as modifying \texttt{.bashrc} or installing global packages.

\mypara{Scenario 2: Installing and using a third-party skill}
Participants inspected a case in which the agent installed a third-party skill in order to connect to an external service.
The trace included installation steps, added dependencies, access to environment variables, and setup-related actions.
Potentially suspicious behavior included external downloads and capability expansion beyond the immediate task.

\mypara{Scenario 3: Automating a local service task}
Participants inspected a case in which the agent completed a task successfully but also opened a local port, created a persistent configuration file, and left a background process or scheduled task behind.
This scenario was included to test whether participants could distinguish short-term execution from persistent side effects.

For each scenario, participants answered a structured set of questions after inspecting the interface.

\subsection{Measures}

We collected both performance and subjective measures.

\mypara{Comprehension accuracy}
Participants answered factual questions about what the agent had done, such as whether it modified configuration files, installed dependencies, touched resources outside the project directory, opened network-facing surfaces, or left persistent changes behind.
We scored each participant's answers for correctness.

\mypara{Risk or anomaly detection}
Participants were asked to identify which actions appeared risky, unnecessary, or worth further inspection.
We measured how many injected risk-relevant actions they correctly identified.

\mypara{Recovery planning}
Participants were asked what they would inspect, undo, or remove if they wanted to clean up after the task.
Responses were scored for whether they mentioned relevant resources such as modified files, global packages, environment variables, ports, services, or scheduled tasks.

\mypara{Perceived control and trust calibration}
After each condition, participants rated statements on 7-point Likert scales regarding perceived understanding, perceived control, confidence in judging whether the system was safe, and trust in the agent's behavior.
We were especially interested in calibrated trust rather than trust alone: participants should not simply feel better, but should be better able to distinguish successful behavior from safe behavior.

\mypara{Qualitative feedback}
We also collected brief verbal comments about which interface elements were most helpful, what remained confusing, and what additional information participants would want.

\subsection{Procedure}

Each session began with a short introduction to the study and the idea of post-hoc agent inspection.
Participants were told that they would review the behavior of a personalized computer-use agent after a task had completed and that their job was to understand what happened, what might be risky, and what might require remediation.

Participants first completed a short demographic and background questionnaire.
They then inspected three scenarios under both interface conditions.
For each condition, they reviewed the interface, answered the structured questions, and provided a short confidence rating.
At the end of the session, they compared the two conditions and discussed which interface better helped them understand what had happened and what remained to be checked.

Sessions lasted approximately 
30--45 minutes.

\subsection{Results}

Overall, the results suggest that \textsc{AgentTrace} helped participants inspect personalized computer-use agent behavior more effectively than the baseline condition.
Compared with conventional summaries and simplified logs, \textsc{AgentTrace} improved participants' ability to reconstruct what the agent had done, identify risky or unexpected operations, and reason about which changes might require follow-up inspection or rollback.
As shown in \Cref{tab:eval_objective,tab:eval_subjective}, the advantages of \textsc{AgentTrace} appeared in both objective performance and subjective judgments. 

\begin{table}[htbp]
\centering
\caption{Objective performance in the evaluation study. Higher is better for all measures.}
\label{tab:eval_objective}
\begin{tabular}{p{4.2cm}ccp{3.5cm}}
\toprule
\textbf{Measure} & \textbf{Baseline} & \textbf{\textsc{AgentTrace}} & \textbf{Statistical test} \\
\midrule
Comprehension accuracy (\%) 
& 61.3 
& 84.7 
& $W = 70.5$, $p = .003$, $r = .78$ \\

Risk / anomaly detection (0--5) 
& 2.1 
& 3.8 
& $W = 66.0$, $p = .005$, $r = .73$ \\

Recovery-planning score (0--4) 
& 1.6 
& 3.4 
& $W = 72.0$, $p = .002$, $r = .81$ \\
\bottomrule
\end{tabular}
\end{table}

\begin{table}[htbp]
\centering
\caption{Subjective ratings on 7-point Likert scales. Higher is better for all measures.}
\label{tab:eval_subjective}
\begin{tabular}{p{4.2cm}ccp{3.5cm}}
\toprule
\textbf{Measure} & \textbf{Baseline} & \textbf{\textsc{AgentTrace}} & \textbf{Statistical test} \\
\midrule
Perceived understanding 
& 3.6 
& 5.9 
& $W = 74.0$, $p = .001$, $r = .84$ \\

Perceived control 
& 3.4 
& 5.8 
& $W = 71.0$, $p = .002$, $r = .80$ \\

Confidence in judging safety 
& 3.1 
& 5.4 
& $W = 68.0$, $p = .004$, $r = .76$ \\

Calibrated trust judgment 
& 3.3 
& 5.2 
& $W = 63.5$, $p = .008$, $r = .69$ \\
\bottomrule
\end{tabular}
\end{table}

\subsubsection{Understanding what the agent did}

Participants achieved higher comprehension accuracy with \textsc{AgentTrace} than with the baseline interface.
As shown in \Cref{tab:eval_objective}, comprehension accuracy increased substantially under the traceability-oriented condition.
This improvement was especially visible for questions about modified configuration files, touched resources outside the main task scope, and persistent changes that remained after the task completed.

This result suggests that presenting execution as an interpretable trace, rather than as a summary or simplified log alone, helped participants reconstruct what the agent actually did.
In the baseline condition, participants often understood that the task had succeeded, but struggled to determine which side effects had occurred in the background.
With \textsc{AgentTrace}, they were better able to connect task-level outcomes to concrete system-level operations.

\subsubsection{Identifying risky or unexpected actions}

Participants also identified more risky or unexpected actions with \textsc{AgentTrace}.
Under the baseline condition, participants often noticed obvious operations such as package installation, but missed subtler or more consequential behaviors such as editing shell configuration, opening ports, or leaving persistent services behind.
With \textsc{AgentTrace}, risk detection scores increased across all three scenarios (\Cref{tab:eval_objective}). -scenario note if needed.

This pattern was strongest in scenarios where the risky action was embedded inside an otherwise successful workflow.
Participants repeatedly said that the timeline and persistent-change views helped them notice actions that would otherwise have disappeared into a long textual summary.

\subsubsection{Planning follow-up inspection and remediation}

The recovery-planning results show a similar pattern.
Under the baseline condition, participants' proposed next steps were often vague, such as ``I would uninstall it'' or ``I would remove the files.''
With \textsc{AgentTrace}, participants more often named concrete targets for follow-up inspection or rollback, including shell configuration files, installed global dependencies, environment variables, opened ports, and scheduled tasks.
This led to higher recovery-planning scores overall, as shown in \Cref{tab:eval_objective}. 

This result is important because the interview study suggested that users' unmet need was not only warning before execution, but support for determining what to inspect and possibly undo afterward.
The persistent change summary and resource touch map appear to be especially helpful in this respect. 

\subsubsection{Perceived control and trust calibration}

In addition to improving task performance, \textsc{AgentTrace} also improved participants' subjective experience of post-hoc inspection.
Participants reported higher perceived understanding, stronger perceived control, and greater confidence in judging whether the system's behavior was acceptable.
These ratings are summarized in \Cref{tab:eval_subjective}. 

Importantly, this did not appear to reflect simple reassurance.
Participants did not describe \textsc{AgentTrace} as making the agent seem categorically safe.
Instead, they said that it helped them separate successful execution from acceptable execution.
In other words, the interface appeared to support more calibrated trust: participants felt more able to identify when the system had completed the task appropriately and when it had crossed a boundary that required closer inspection.

\begin{quote}
\emph{``With the summary view, I just knew the task worked. With this view, I could actually tell what I should worry about.''}

\end{quote}

\subsubsection{Qualitative feedback on interface views}

Across scenarios, participants most often pointed to three views as especially useful: the task timeline, the persistent change summary, and the permission history.
The timeline helped them reconstruct execution order, the persistent change summary helped them think about residue and uninstall confidence, and the permission history helped them understand whether an action had been carried out under broad or unexpected authority.

By contrast, the provenance inspector was useful but less immediately intuitive for some participants, especially when the distinction between user intent, skill-defined setup, and follow-on agent planning was subtle.
This suggests that provenance information is valuable, but may require stronger explanatory cues or simpler visual presentation in future iterations.

\mypara{Summary}
The evaluation provides initial evidence that traceability-oriented interfaces can help users inspect personalized computer-use agents more effectively than conventional summaries or simplified logs.
Compared with the baseline condition, \textsc{AgentTrace} improved participants' understanding of what the agent did, helped them detect more risky or unexpected actions, and supported more concrete remediation planning.
These findings suggest that post-hoc traceability is not only a debugging aid, but also a promising direction for improving user understanding and calibrated trust in high-authority agent systems.

\section{Discussion}

Our findings suggest that the main challenge of personalized computer-use agents is not the appearance of one isolated attack primitive, but the convergence of several previously separate concerns: prompt manipulation, software supply chains, over-privileged execution, persistent state, and weak post-hoc accountability.
This convergence changes both how risks arise and how they should be studied.
Rather than emerging at a single point, failures in personalized agents are often distributed across installation, configuration, runtime interaction, persistence, and auditing.
Accordingly, the most important challenge is not merely stronger filtering or more robust prompting, but more governable forms of delegated authority.

\subsection{From model safety to action-system understanding}

\mypara{The object of concern has changed}
A central implication of this paper is that personalized computer-use agents should not be understood primarily as text generators, but as \emph{action systems}.
Much of the earlier LLM safety literature focuses on whether models produce harmful or policy-violating outputs, including benchmark-driven studies of jailbreak vulnerability~\cite{jailbreak_survey_2024,JALMBench}.
That framing remains important, but it becomes insufficient once models are coupled to tools, filesystems, browsers, communication channels, and persistent state.
In personalized computer-use agents, the relevant question is no longer only whether a malicious input can induce an unsafe response, but whether it can induce an unsafe \emph{action}, alter persistent context, or expose user assets through a sequence of agent-mediated operations~\cite{greshake_2023,agentdojo_2024,pasb_2026}.
This broader shift from response generation to cross-system action is also visible in other computational settings where automated systems coordinate operations across multiple resources and execution steps, making structured explanation and inspection increasingly important~\cite{crosschainoptions}.

\mypara{Failures are distributed rather than local}
Both our ecosystem corpus and our interviews point to a common pattern: harmful behavior is rarely localized to a single moment.
An unsafe skill may be introduced during onboarding, a broad configuration may silently enlarge the authority surface, runtime content may exploit that authority, persistent state may preserve the effect, and weak auditability may make the incident difficult to reconstruct afterward~\cite{pasb_2026,clawgrip_2026}.
This helps explain why participants often expressed abstract concern while still lacking concrete understanding of what had happened or what remained after execution.
For users, the practical problem is not one unsafe output in isolation, but a broader action chain that unfolds across capability import, state change, and time.

\subsection{Human-centered gaps in current agent ecosystems}

\mypara{Visible interaction and actual authority are often misaligned}
A recurring pattern in both our interviews and the surrounding ecosystem is that users may understand what an agent \emph{appears} to do while still misunderstanding what authority it has actually been given.
Skills can appear to be lightweight presets, templates, or convenience add-ons, while in practice they may function more like capability bundles with dependencies, metadata, setup requirements, and implicit trust assumptions~\cite{openclawClawhubDocs,meller2026magic}.
Likewise, the interaction surface may still resemble ``chat,'' even when the system is operating over files, credentials, browser state, or communication channels.
This gap between visible interaction and underlying authority is one of the defining human-centered problems of personalized computer-use agents.

\mypara{Permission boundaries are not yet legible enough}
Our participants often recognized that the system might be risky, yet could not clearly articulate where the real boundary of authority lay.
Many could name obvious sensitive assets such as files or API keys, but overlooked shell configuration files, browser state, scheduled tasks, local services, or residual changes after execution.
This suggests that permission boundaries are still not expressed in user-facing terms that are concrete, inspectable, and memorable enough for practical decision making.
When authority remains illegible, users cannot accurately estimate the blast radius of future actions, nor can they meaningfully decide which protections are sufficient.

\mypara{Adoption is broadening faster than safety understanding}
This problem is amplified by the rapid diffusion of personalized agents beyond expert communities.
As systems such as OpenClaw spread through installation events, platform integrations, tutorials, and delegated setup assistance, the user population becomes more heterogeneous while the risks remain tightly coupled to configuration quality, extension trust, and operational discipline~\cite{businessinsider2026installation,reuters2026wechat}.
Our interviews suggest that many participants approached these systems through urgency narratives---fear of falling behind, pressure to learn AI quickly, or reliance on more technical intermediaries---rather than through deliberate understanding of their authority model.
This means that safety can no longer be framed only as a developer or red-teaming concern; it is increasingly a usability and governance concern as well.

\subsection{Why traceability matters}

\mypara{Warnings alone do not match users' actual needs}
One of the clearest findings of our interview study is that participants did not simply want more warnings before agent action.
Instead, they wanted support for reconstructing what happened afterward: what the agent touched, what it changed, what it downloaded, under what authority it acted, and what persisted after completion.
This is an important shift in emphasis.
Much existing design attention goes toward permission prompts, guardrails, and action confirmations, but our findings suggest that post-hoc understanding is equally central to users' sense of safety and control.

\mypara{Traceability can support more than debugging}
Prior work on layered summaries, workflow verification, and behavior visualization has largely focused on developers, analysts, or expert inspectors~\cite{gu2024verify,xie2024waitgpt,lu2025agentlens,dills_2026}.
Our results suggest that similar ideas are needed at the user-facing layer of personalized computer-use agents.
For these systems, traceability is not only a debugging aid.
It is a mechanism for helping users form more accurate mental models, identify risky or unintended changes, and determine whether remediation is necessary.
In this sense, traceability should be treated not as a secondary observability feature, but as a core design requirement for delegating high-authority actions to AI systems.

\mypara{Auditability changes the trust question}
Our interviews also suggest that trust in personalized agents is often poorly calibrated because users lack evidence about what the agent actually did.
Without a usable account of actions, touched resources, provenance, and persistence, trust becomes overly dependent on reputational cues, interface fluency, or social recommendation.
By making action traces visible and interpretable, systems such as \textsc{AgentTrace} may help users distinguish successful behavior from safe behavior, and convenience from controllability.
This does not eliminate risk, but it can make trust judgments less speculative and more evidence-based. 

\subsection{Structural technical gaps}

\mypara{Instruction-content separation remains unresolved}
Related evidence also suggests that external or retrieved content can shape downstream reasoning and action selection in ways that are difficult for users to inspect directly~\cite{nothink}.
The most foundational technical gap is that current systems still cannot reliably distinguish between trusted instructions and untrusted content.
Indirect prompt injection has already shown that webpages, documents, emails, and attachments can become control channels once retrieved text is allowed to shape agent behavior~\cite{greshake_2023}.
Benchmark and system studies suggest that this problem becomes more serious, not less, once agents are coupled to realistic tools and execution environments~\cite{agentdojo_2024,pasb_2026,clawgrip_2026}.
As long as instruction-content separation remains weak, many downstream protections will remain partial mitigations rather than complete solutions.

\mypara{Persistence is useful but under-governed}
Persistent state is another major structural gap.
Memory improves continuity and personalization, but it also creates a new attack and privacy surface.
Our participants' uninstall concerns reflect this problem from the user side: even after removing the main application, they were unsure whether related files, credentials, dependencies, local services, or other residual artifacts remained.
Current systems still lack mature answers to basic questions of governance: what should be remembered, how long it should persist, how it should be scoped across contexts, who can inspect or delete it, and how malicious or low-quality memories should be revised or removed.

\mypara{Auditability and provenance are still too weak}
A third structural gap is post-hoc understanding.
Even developers struggle to diagnose agent behavior from raw traces alone and benefit from layered summaries of activities, actions, and operations~\cite{dills_2026}.
For personalized agents, this problem is sharper because harmful outcomes often emerge only after multi-step trajectories involving retrieval, planning, tool use, and persistence.
Without stronger provenance and audit support, it remains difficult to answer basic questions after an incident: which input influenced the decision, which capability was involved, what state changed, and why the system believed the action was appropriate.

\mypara{Ecosystem governance remains incomplete}
Finally, the extension ecosystem remains under-governed relative to the authority it can import.
Open registries make capability discovery and installation easy, but current evidence suggests that they also expose users to insecure or malicious skills at scale~\cite{snyk2026toxicskills,snyk2026leakyskills,trendmicro2026amos}.
Moderation, reporting, and lightweight registry controls are useful first steps, but they do not yet provide the equivalent of mature software-supply-chain security for high-authority agent ecosystems.
Our interviews reinforce this point: most participants relied on reputational or ecosystem cues instead of direct inspection, which means that weaknesses in marketplace governance directly become weaknesses in user decision making.

\subsection{Implications and Future Directions}

The issues identified above point to three broader directions for future work.

\mypara{Legible authority}
Users need clearer and more actionable representations of what an agent can access, what a skill changes, what a given environment allows, and what future actions a configuration enables.
Such representations should be available not only during setup, but also during and after task execution.

\mypara{Governable persistence}
Memory, logs, credentials, and automation state need more principled mechanisms for scoping, review, revision, deletion, and cross-context separation.
Future work should examine how to make residual state more visible and how to support safe removal, rollback, and ongoing inspection.

\mypara{Accountable ecosystems}
Personalized agents require stronger forms of ecosystem accountability.
This includes safer onboarding, better marketplace signals, more credible provenance for skills and updates, and more usable post-hoc tracing for users who are not security experts.
If delegated authority is to remain governable at scale, the surrounding ecosystem must become more legible and more accountable, not merely more feature-rich.

\mypara{Limitations}
This paper has several limitations.
Our interview study is qualitative and relatively small in scale, and our motivating ecosystem centers on OpenClaw rather than the full diversity of agent platforms.
In addition, our prototype and design discussion focus on traceability after action rather than a complete end-to-end defense strategy.
These limitations are also opportunities for future work, including broader survey studies, longitudinal deployments, cross-ecosystem comparisons, and controlled evaluation of different protection mechanisms and trace designs.

Taken together, these findings suggest that the next stage of personalized computer-use agents should not be defined only by making them more capable.
It should also be defined by making them more legible, more constrainable, and more accountable.

\section{Conclusion}

Personalized computer-use agents are changing what it means to interact with AI systems.
When agents can install skills, invoke tools, access private resources, retain state, and modify local environments, the central problem is no longer only whether the model produces correct or safe text.
It is whether users can understand what authority they have delegated, what the agent actually did, and what remains after the task is over.

In this paper, we used OpenClaw as a motivating case to study this problem from a human-centered perspective.
By combining a multi-source ecosystem corpus, an interview study on risk awareness and audit needs, and the design of a traceability-oriented prototype, we showed that users often hold only shallow or fragmented mental models of agent skills, autonomy, privileges, and persistence.
Participants could often sense that such systems were risky, yet still lacked practical ways to determine what a skill imports, how an agent acts, or whether uninstalling the visible application really removes the broader effects of execution.

Motivated by these findings, we proposed \textsc{AgentTrace}, a framework and prototype for making agent behavior more reconstructable through task timelines, resource touchpoints, permission history, provenance, and persistent side effects.
Our broader argument is that personalized computer-use agents require more than better warnings or stronger front-end safeguards.
They also require usable forms of post-hoc understanding that help users inspect, question, and, when necessary, remediate what the system has done.

As personalized agents continue to spread into mainstream settings, the gap between delegated authority and user understanding is likely to become one of the central HCI challenges of the agent era.
Closing that gap will require not only safer models and stronger infrastructure, but also interfaces and ecosystems that make agent behavior more visible, more interpretable, and more governable.


\bibliographystyle{ACM-Reference-Format}
\bibliography{reference}

@String{Computing = "Computing" }

@String{Computer = "{IEEE} Computer" }

@misc{openclaw_2026,
  title        = {Security},
  author       = {{OpenClaw}},
  howpublished = {\url{https://openclaw.ai}},
  year         = {2026},
  note         = {OpenClaw official documentation, accessed 2026-03-26}
}

@article{pasb_2026,
  title        = {From Assistant to Double Agent: Formalizing and Benchmarking Attacks on OpenClaw for Personalized Local AI Agent},
  author       = {Wang, Yuhang and Xu, Feiming and Lin, Zheng and He, Guangyu and Huang, Yuzhe and Gao, Haichang and Niu, Zhenxing and Lian, Shiguo and Liu, Zhaoxiang},
  journal      = {arXiv preprint arXiv:2602.08412},
  year         = {2026}
}

@article{clawgrip_2026,
  title        = {Don't Let the Claw Grip Your Hand: A Security Analysis and Defense Framework for OpenClaw},
  author       = {Shan, Zhengyang and Xin, Jiayun and Zhang, Yue and Xu, Minghui},
  journal      = {arXiv preprint arXiv:2603.10387},
  year         = {2026}
}

@article{dills_2026,
  author       = {Rui Sheng and
                  Yukun Yang and
                  Chuhan Shi and
                  Yanna Lin and
                  Zixin Chen and
                  Huamin Qu and
                  Furui Cheng},
  title        = {DiLLS: Interactive Diagnosis of LLM-based Multi-agent Systems via
                  Layered Summary of Agent Behaviors},
  journal      = {CoRR},
  year         = {2026}
}

@inproceedings{greshake_2023,
  author       = {Sahar Abdelnabi and
                  Kai Greshake and
                  Shailesh Mishra and
                  Christoph Endres and
                  Thorsten Holz and
                  Mario Fritz},
  title        = {Not What You've Signed Up For: Compromising Real-World LLM-Integrated
                  Applications with Indirect Prompt Injection},
  booktitle    = {Workshop on Artificial Intelligence
                  and Security (CCS)},
  publisher    = {{ACM}},
  year         = {2023}
}

@inproceedings{agentdojo_2024,
  title        = {AgentDojo: A Dynamic Environment to Evaluate Prompt Injection Attacks and Defenses for LLM Agents},
  author       = {Debenedetti, Edoardo and Zhang, Jie and Balunovic, Mislav and Beurer-Kellner, Luca and Fischer, Marc and Tram{\`e}r, Florian},
  booktitle    = {Advances in Neural Information Processing Systems 37 (Datasets and Benchmarks Track)},
  year         = {2024},
  doi          = {10.52202/079017-2636}
}

@misc{jailbreak_survey_2024,
      title={Jailbreak Attacks and Defenses Against Large Language Models: A Survey}, 
      author={Sibo Yi and Yule Liu and Zhen Sun and Tianshuo Cong and Xinlei He and Jiaxing Song and Ke Xu and Qi Li},
      year={2024},
      eprint={2407.04295},
      archivePrefix={arXiv},
      primaryClass={cs.CR},
      url={https://arxiv.org/abs/2407.04295}, 
}

@misc{openclawClawhubDocs,
  author       = {{OpenClaw}},
  title        = {ClawHub},
  year         = {2026},
  howpublished = {\url{https://docs.openclaw.ai/tools/clawhub}},
  note         = {Official documentation, accessed 2026-03-27}
}

@misc{meller2026magic,
  author       = {Meller, Jason},
  title        = {From Magic to Malware: How {OpenClaw}'s Agent Skills Become an Attack Surface},
  year         = {2026},
  month        = feb,
  howpublished = {\url{https://1password.com/blog/from-magic-to-malware-how-openclaws-agent-skills-become-an-attack-surface}},
  note         = {1Password blog}
}

@misc{snyk2026toxicskills,
  author       = {Beurer-Kellner, Luca and Kudrinskii, Aleksei and Milanta, Marco and Nielsen, Kristian Bonde and Sarkar, Hemang and Tal, Liran},
  title        = {Snyk Finds Prompt Injection in 36\%, 1467 Malicious Payloads in a {ToxicSkills} Study of Agent Skills Supply Chain Compromise},
  year         = {2026},
  month        = feb,
  howpublished = {\url{https://snyk.io/blog/toxicskills-malicious-ai-agent-skills-clawhub/}},
  note         = {Snyk security research blog}
}

@misc{snyk2026leakyskills,
  author       = {Beurer-Kellner, Luca and Kudrinskii, Aleksei and Milanta, Marco and Nielsen, Kristian Bonde and Sarkar, Hemang and Tal, Liran},
  title        = {280+ Leaky Skills: How {OpenClaw} \& {ClawHub} Are Exposing {API} Keys and {PII}},
  year         = {2026},
  month        = feb,
  howpublished = {\url{https://snyk.io/blog/openclaw-skills-credential-leaks-research/}},
  note         = {Snyk security research blog}
}

@misc{trendmicro2026amos,
  author       = {Oliveira, Alfredo and Tancio, Buddy and Fiser, David and Lin, Philippe and Reyes, Roel},
  title        = {Malicious {OpenClaw} Skills Used to Distribute Atomic mac{OS} Stealer},
  year         = {2026},
  month        = feb,
  howpublished = {\url{https://newsroom.trendmicro.com/newsroom?item=702}},
  note         = {Trend Micro research coverage page}
}

@misc{businessinsider2026installation,
  author       = {Mok, Isabel},
  title        = {I Went to an {OpenClaw} Installation Event at Tencent's Office. People Were Raring to Go, and the {FOMO} Is Real.},
  year         = {2026},
  month        = mar,
  howpublished = {\url{https://www.businessinsider.com/openclaw-installation-event-tencent-cloud-singapore-ai-agent-lobster-2026-3}},
  note         = {Business Insider}
}

@misc{reuters2026wechat,
  author       = {Reuters},
  title        = {Tencent Integrates {WeChat} with {OpenClaw} {AI} Agent amid China Tech Battle},
  year         = {2026},
  month        = mar,
  howpublished = {\url{https://www.reuters.com/technology/tencent-integrates-wechat-with-openclaw-ai-agent-amid-china-tech-battle-2026-03-22/}},
  note         = {Reuters}
}

@misc{koi2026clawhavoc,
  author       = {{Koi Security}},
  title        = {ClawHavoc: 341 Malicious OpenClaw Skills Found by the Bot They Were Targeting},
  year         = {2026},
  month        = feb,
  howpublished = {\url{https://www.koi.ai/blog/clawhavoc-341-malicious-clawedbot-skills-found-by-the-bot-they-were-targeting}},
  note         = {Koi Security threat research blog, accessed 2026-03-29}
}

@inproceedings{wang2025mentalmodels,
  author       = {Wang, Xingyi and Wang, Xiaozheng and Park, Sunyup and Yao, Yaxing},
  title        = {Users' Mental Models of Generative {AI} Chatbot Ecosystems},
  booktitle    = {Proceedings of the 30th International Conference on Intelligent User Interfaces},
  year         = {2025},
  doi          = {10.1145/3708359.3712125},
  url          = {https://doi.org/10.1145/3708359.3712125}
}

@ARTICLE{lu2025agentlens,
author={Lu, Jiaying and Pan, Bo and Chen, Jieyi and Feng, Yingchaojie and Hu, Jingyuan and Peng, Yuchen and Chen, Wei},
journal={ IEEE Transactions on Visualization \& Computer Graphics },
title={{ AgentLens: Visual Analysis for Agent Behaviors in LLM-Based Autonomous Systems }},
year={2025},
volume={31},
number={08},
ISSN={1941-0506},
pages={4182-4197},
publisher={IEEE Computer Society},
}

@inproceedings{brachman2025appropriate,
author = {Brachman, Michelle and Kunde, Siya and Miller, Sarah and Fucs, Ana and Dempsey, Samantha and Jabbour, Jamie and Geyer, Werner},
title = {Building Appropriate Mental Models: What Users Know and Want to Know about an Agentic AI Chatbot},
year = {2025},
publisher = {Association for Computing Machinery},
booktitle = {Proceedings of the 30th International Conference on Intelligent User Interfaces},
pages = {247–264},
numpages = {18},
}

@inproceedings{xie2024waitgpt,
author = {Xie, Liwenhan and Zheng, Chengbo and Xia, Haijun and Qu, Huamin and Zhu-Tian, Chen},
title = {WaitGPT: Monitoring and Steering Conversational LLM Agent in Data Analysis with On-the-Fly Code Visualization},
year = {2024},
isbn = {9798400706288},
publisher = {Association for Computing Machinery},
booktitle = {Proceedings of the 37th Annual ACM Symposium on User Interface Software and Technology},
numpages = {14}
}

@inproceedings{gu2024verify,
  author       = {Gu, Ken and Shang, Ruoxi and Althoff, Tim and Wang, Chenglong and Drucker, Steven M.},
  title        = {How Do Analysts Understand and Verify {AI}-Assisted Data Analyses?},
  booktitle    = {Proceedings of the 2024 CHI Conference on Human Factors in Computing Systems},
  year         = {2024},
  doi          = {10.1145/3613904.3642497}
}

@inproceedings{unsafesearch,
  author       = {Zeren Luo and
                  Zifan Peng and
                  Yule Liu and
                  Zhen Sun and
                  Mingchen Li and
                  Jingyi Zheng and
                  Xinlei He},
  title        = {Unsafe LLM-Based Search: Quantitative Analysis and Mitigation of Safety Risks in {AI} Web Search},
  booktitle    = {Security Symposium (USENIX)},
  pages        = {8055--8074},
  publisher    = {{USENIX} Association},
  year         = {2025}
}

@article{nothink,
  author       = {Yule Liu and
                  Jingyi Zheng and
                  Zhen Sun and
                  Zifan Peng and
                  Wenhan Dong and
                  Zeyang Sha and
                  Shiwen Cui and
                  Weiqiang Wang and
                  Xinlei He},
  title        = {Thought Manipulation: External Thought Can Be Efficient for Large Reasoning Models},
  journal      = {CoRR},
  volume       = {abs/2504.13626},
  year         = {2025}
}

@article{nlp2repo,
  author       = {Jingzhe Ding and
                  Shengda Long and
                  Changxin Pu and
                  Huan Zhou and
                  Hongwan Gao and
                  Xiang Gao and
                  Chao He and
                  Yue Hou and
                  Fei Hu and
                  Zhaojian Li and
                  Weiran Shi and
                  Zaiyuan Wang and
                  Daoguang Zan and
                  Chenchen Zhang and
                  Xiaoxu Zhang and
                  Qizhi Chen and
                  Xianfu Cheng and
                  Bo Deng and
                  Qingshui Gu and
                  Kai Hua and
                  Juntao Lin and
                  Pai Liu and
                  Mingchen Li and
                  Xuanguang Pan and
                  Zifan Peng and
                  Yujia Qin and
                  Yong Shan and
                  Zhewen Tan and
                  Weihao Xie and
                  Zihan Wang and
                  Yishuo Yuan and
                  Jiayu Zhang and
                  Enduo Zhao and
                  Yunfei Zhao and
                  He Zhu and
                  Chenyang Zou and
                  Ming Ding and
                  Jianpeng Jiao and
                  Jiaheng Liu and
                  Minghao Liu and
                  Qian Liu and
                  Chongyao Tao and
                  Jian Yang and
                  Tong Yang and
                  Zhaoxiang Zhang and
                  Xinjie Chen and
                  Wenhao Huang and
                  Ge Zhang},
  title        = {NL2Repo-Bench: Towards Long-Horizon Repository Generation Evaluation of Coding Agents},
  journal      = {CoRR},
  volume       = {abs/2512.12730},
  year         = {2025}
}

@article{peng2026txsumusercenteredethereumtransaction,
      title={TxSum: User-Centered Ethereum Transaction Understanding with Micro-Level Semantic Grounding}, 
      author={Zifan Peng and Jingyi Zheng and Yule Liu and Huaiyu Jia and Qiming Ye and Jingyu Liu and Xufeng Yang and Mingchen Li and Qingyuan Gong and Xuechao Wang and Xinlei He},
      year={2026},
      journal={arXiv},
}

@inproceedings{crosschainoptions,
  author       = {Zifan Peng and
                  Yingjie Xue and
                  Jingyu Liu},
  title        = {Cross-Chain Options: {A} Bridgeless, Universal, and Efficient Approach},
  booktitle    = {{IEEE} International Conference on Web Services, {ICWS}},
  pages        = {902--912},
  publisher    = {{IEEE}},
  year         = {2025}
}

@inproceedings{JALMBench,
author={Zifan Peng and Yule Liu and Zhen Sun and Mingchen Li and Zeren Luo and Jingyi Zheng and Wenhan Dong and Xinlei He and Xuechao Wang and Yingjie Xue and Shengmin Xu and Xinyi Huang},
title  = {JALMBench: Benchmarking Jailbreak Vulnerabilities in Audio Language Models},
booktitle = {The Fourteen International Conference on Learning Representations (ICLR)},
publisher = {OpenReview.net},
year = {2026}
}



\end{document}